\documentclass[fleqn,usenatbib]{mnras}

\usepackage{newtxtext,newtxmath}

\usepackage[T1]{fontenc}
\usepackage{soul}
\setstcolor{red}
\DeclareRobustCommand{\VAN}[3]{#2}
\let\VANthebibliography\thebibliography
\def\thebibliography{\DeclareRobustCommand{\VAN}[3]{##3}\VANthebibliography}

\usepackage{graphicx}	
\usepackage{amsmath}	
\usepackage{amssymb}	
\usepackage[flushleft]{threeparttable}
\usepackage[usenames]{color}

\title[Two-wind interactions]{Two-wind interactions in binaries with two orbiting giant stars}

\author[Castellanos-Ram\'\i rez et al.]{
A. Castellanos-Ram\'irez,$^{1}$\thanks{E-mail: acastellanos@astro.unam.mx}
A. Rodr\'iguez-Gonz\'alez,$^{2}$
Z. Meliani$^{3}$
P.R. Rivera-Ortiz$^{4}$
A.C. Raga$^{2}$
and J. Cant\'o$^{1}$
\\
$^{1}$ Universidad Nacional Aut\'onoma de M\'exico, Instituto de Astronom\'\i a, Ap. 70-264, CDMX, 04510, M\'exico\\
$^{2}$Instituto de Ciencias Nucleares, Universidad Nacional Aut\'onoma de M\'exico, Ap. 70-543, 04510 CDMX, M\'exico\\
$^{3}$LUTH, Observatoire de Paris, CNRS, PSL, Universit\'{e} de Paris; 5 Place Jules Janssen, 92190 Meudon, France\\
$^{4}$Univ. Grenoble Alpes, CNRS, IPAG, 38000 Grenoble, France
}

\date{Accepted XXX. Received YYY; in original form ZZZ}

\pubyear{2021}

\begin{document}
\label{firstpage}
\pagerange{\pageref{firstpage}--\pageref{lastpage}}
\maketitle
%
\begin{abstract}
Some red giant envelopes present spiral structures (seen either in dust scattered stellar continuum or in molecular line emission), the most striking example probably being AFGL 3068. This object has been modeled (both analytically and numerically) in terms of a wind ejected from a star in orbit around a binary companion. We revisit both analytical models and 3D simulations of a wind from an orbiting red giant star, and extend the numerical simulations to the case of a binary with two red giants with strong winds.
We find that most two-wind models on the orbital plane show a ``double spiral'' structure close to the binary source, and that these two arms merge into a single spiral structure at larger distances. However, for the case of a binary with two identical winds the two spiral arms are still present at large distances from the binary source.
We also find that for models of two (not identical) dynamically important winds, a region close to the orbital plane has material from both winds. Also, an approximately conical region centered on the orbital axis is filled exclusively by the wind with larger momentum rate. These two structures lead to morphologies reminiscent of the so-called ``hour glass'' planetary nebulae.
Finally, we find that increasing wind velocity disparities lead to the formation of clumpy structures along the spiral amrs. Observations of ``clumpy spirals'' are therefore likely to indicate the presence of two strong winds from the stars in the central binary system.
\end{abstract}




\section{Introduction}
The diverse morphology of planetary nebulae (bipolar, multipolar, elliptical, etc.) contrasts with that of their progenitors, low and intermediate mass stars in the Asymptotic Giant Branch (AGB), which have almost spherical shapes (see, e.g., \citealt{Mauron2006}). It seems that, although an individual AGB star has the possibility of forming morphologies away from spherical symmetry, it does not have enough energy or angular momentum to develop high asymmetries. It also cannot maintain the magnetic field strength for the time required to form a bipolar-type PNe (see the brief discussion in the third section of \citealt{Lagadec2015} and the references therein). Interesting examples of non-spherically symmetric AGB winds are given by \cite{Freytag2017}, who performed 3D radiation-hydrodynamics simulations of individual AGB stars. These simulations produce winds with high density contrast structures, which do not show bipolar/multipolar symmetries nor jet-like features.


The combination of a latitude-dependent AGB wind with a later, isotropic fast wind from a PNe central star can result in the production of bipolar nebulae (see, e.g., \citealt{Frank1993}). However, there is currently no viable
model for the strong equatorial enhancement in the AGB wind that is required. Bipolar nebulae can also be produced with a wind that has a strong enough toroidal magnetic field (see, e.g., \citealt{Garcia-Segura2005}). The more extreme departures from spherical symmetry seen in multipolar PNe defiy an interpretation in terms of winds from a single stellar source.

There is now a lot of evidence that the marked difference in symmetry between the progenitors and PNe may be due to the interaction of the AGB progenitor with a binary companion
(e.g., \citealt{Homan2020}).
In the case of a close binary (with separations $\sim 1$~AU),
when one of the stars grows to become a red giant, the second
star is engulfed by the expanding component. This results
in a ``common envelope'' phase with complex properties (see,
e.g., \citealt{Soker1998} and the reviews of \citealt{Ivanova2013} and \citealt{Lagadec2015}). 

In the case of detached (wide) binaries such as AFGL 3068
\citep{Mauron2006},  the interaction between the two stars only occurs through their stellar winds throughout the evolution of the binary  \citep{Mastrodemos1998}. In the present paper, we will focus on this type of two-wind interaction.

The problem of an isotropic wind ejected from a star in a wide binary system with a circular orbit was first studied by \citet{Soker1994}. This work shows that the orbital motion results in the production of spiral shock structures in the orbital plane, which form as a result of the enhanced velocity (in the center of mass frame of the binary) of the outflow in the instantaneous direction of the orbital motion of the stellar wind source.

\citet{Mastrodemos1999}  used 3D smoothed particle hydrodynamic (SPH) to investigate the dynamics of an initially spherical, dusty wind from an AGB star in orbit around a binary companion (exploring different masses for the companion, binary separation distances and wind velocities). On the orbital plane, they obtain the spiral shocks of the analytic model of \citet{Soker1994}, as well as extensions of these shocks away from the orbital plane. This extension of the spiral shocks to higher latitudes was studied analytically by \citet{Canto1999}, who find that in cuts perpendicular to the orbital plane the shock structure has a morphology of circular arcs that do not reach the orbital axis, leaving a cylindrical, axial ``hole'' with no shocks
(this hole is also seen in the simulations of \citealt{Mastrodemos1999}). \citet{Raga2011} showed a successful comparison between the analytic shock structure of \citet{Canto1999} and (Eulerian) 3D gasdynamic simulations, and also explored cases with elliptical binary orbits.

\citet{He2007} studied the spiral pattern produced by wind sources in elliptical orbits with a cold, ``sticky particle'' description, leading to shorter computations than a full gasdynamic description. He presents column density maps obtained for a range of orbital and wind parameters which, depending on the eccentricity and the orientation of the orbital axis with respect to the line of sight, lead to a range of different morphologies. \citet{He2007} shows that the spiral pattern may become broken for eccentricities $e>0.4$.

\citet{Kim2012a} presented 3D gasdynamic simulations exploring the effect of the ratio between the orbital and wind speeds. \citet{Kim2017b} considered eccentric orbits and
found that the binary interaction between AGB stars and a close companion can lead to the formation of circumstellar disks, which might produce a bipolar outflow \citep{Nordhaus2006}. This phenomenon, observed in some PPNe (eg., L2 Pup \citealt{Kervella2016}), was first modelled by \citet{Morris1981}.

From the observational point of view, there are objects that
apparently show the effect of an orbital motion of the source
on the circumstellar environment of AGB stars:
\begin{itemize}
\item \citet{Mauron2006} presented HST images of AFGL 3068, showing a spiral pattern which they interpreted as the shock driven into a red giant wind as a result of an orbital motion of the wind source. These authors found that the AFGL 3068 structure is well fitted by an Archimedean spiral. \citet{Kim2012b} presented detailed (analytic and numerical) models for this object. This result is also supported by ALMA observations of the $^{12}$CO J=$2-1$, $^{13}$CO J=$2-1$
  and HC$_{2}$N J=$24-23$ molecular lines \citep{Kim2017b},
\item \citet{Maercker2012} present CO J=$3-2$ Atacama Large Millimeter Array (ALMA) line maps of the R~Sculptoris structure. These observations, supported by SPH simulations, show the presence of a spiral pattern as a consequence of the orbital motion of the outflow source in a wide binary, 
\item \citet{Claussen2011} present a HC$_3$N J=$4-3$ line map obtained with the Expanded Very Large Array (EVLA) of RW LMi (CIT6), which shows partial rings that do not form a clear spiral structure. \citet{Kim2013} propose that these partial rings can be explained by a motion of the wind source in an eccentric orbit, 
\item \citet{Decin2015} report CO J = $6-5 $ ALMA observations of CW Leo. They use interpretive models (based on the dynamical model of \citealt{Kim2012b}) to model the observed spiral structure,
\item \citet{Harpaz1997} study the protoplanetary nebula CRL 2688 using eccentric orbit models to describe the formation of multiple concentric layers. This result was supported by \citet{Corradi2004} and \citet{Ramos-Larios2016}, who detect ring and arc morphologies in the halos of planetary nebulae, 
\item \citet{Homan2020} report CO, SiO and HCN emission maps of $\pi^1$ Gru obtained with ALMA. They reveal the existence of a spiral pattern and companion stars at 440 AU (a detached companion) and 6 AU (a close companion), 
\item \cite{Kervella2016} report ALMA observations of L$_{2}$ Pup at $\nu=350$GHz.  They find the existence of an equatorial disk and a bipolar flow associated with the presence of a companion massive planet or low-mass brown dwarf.
\end{itemize}

In all of these cases, the morphology of the interaction of
the wind of a primary star with the orbital movement due to
a binary companion (with or without a wind) consists of
a three-dimensional spiral formed by thin layers
of colliding material.

For the case of circular orbits, the shocked structure resembles an Archimedean spiral pattern with a more or less constant spacing between the loops \citep{Kim2017a}.

In this paper, we analyze the case of a wide binary with components of similar masses, in which the giant branch/AGB phases of both stars have a temporal coincidence. The {\it a priori} probability of having a mass ratio $q\sim 1$ in a binary star is probably not very high (see, e.g., \citealt{Hogeveen1992}), so that one is not likely to encounter many binaries with two giant components. However, the masses estimated for the two components of AFGL~3068
(\citealt{Mauron2006} and \citealt{Kim2012b}) are indeed
very similar, so that at least for this object it is possible
that we are observing a binary with two giant stars.

Such a binary will have a spiral structure caused by
the interaction of two winds of comparable mass loss and
momentum rates, combined with the effect of the orbital
motion of the two wind sources. This situation is similar
to the one found in the so-called ``pinwheel nebulae'',
which correspond to spiral wind interaction regions in
binaries with a massive, Wolf-Rayet component (see, e.g.,
\citealt{Tuthill1999} and \citealp{Monnier1999, Monnier2002}).
These systems have been modeled with 2 and 3D hydrodynamical
simulations by \citet{Parkin2008}, \citet{Parkin2009} (for $\eta$ Carinae), \citet{Pittard2009, Pittard2010} (for an O +O system), \citet{vanMarle_etal_2011A&A...527A...3V, Lamberts2012, Hendrix2016}. We study a similar situation, but with much slower winds corresponding to giant stars.

This paper is organized as follows. In section 2 we make a brief review of the analytical model for the case of a single wind from a star in a binary system. In section 3 we present the numerical setup of our models. In section 4, we present the results of the numerical integrations. Finally, our work is summarized in section 5.

\section{The single wind analytical model}

In the case of a circular orbit binary with one of its components emitting an isotropic wind, one can find analytically the structure of the spiral shock in a straightforward way (see the work of \citealt{Canto1999} and \citealt{Raga2011}). Here we summarize this analytic model.

Let $v_w$ be the terminal velocity of the stellar wind, $r_{o}$ the orbital radius, $v_{o}$ the orbital velocity, $P$ the period of the orbit, and $\omega=2\pi/P$ the angular velocity of the stellar wind source. Then, as a consequence of the orbital motion, at a given time $t$,  a rotating spiral shock structure (formed by two, contiguous ``working surface'' style shocks) is formed on the orbital plane (see Figure. \ref{fig:Fig1}).

The locus of this spiral pattern is given by:
\begin{equation}
r_{ws}\approx-\frac{v_w}{\omega}\left[\omega \;\; (\tau-t)+\pi\right],
\label{eq:arquimedes}  
\end{equation}
where $t$ is the present time and $\tau$ is the time at which the material was ejected (satisfying the $\tau\leq t$ condition). The Archimedean spiral given by equation (\ref{eq:arquimedes}) starts at a radius:
\begin{equation}
R_0\approx\frac{v^2_w}{\omega v_o},
\label{eq:startrad}  
\end{equation}
and the spacing between the spiral arms is
\begin{equation}
\Delta r_{ws}=2 \pi \frac{v_w}{\omega}.
\label{eq:Dr}  
\end{equation}
This spiral pattern rotates with angular velocity $\omega$ in the observer's frame of reference. We should note that
equation (\ref{eq:arquimedes}) is valid in the $R_0\gg r_o$ limit (see equation \ref{eq:startrad}), and other terms have
to be included if this condition is not satisfied (see \citealt{Canto1999}).

\cite{Raga2011} show that a spiral pattern is also seen in all planes parallel to the orbital plane. In all of the planes (perpendicular to the orbital plane) that include the orbital axis, the shock structure consists of circular arcs that start on the orbital plane and do not reach the orbital axis, being interrupted at a distance $R_0$ (see equation \ref{eq:startrad}) from this axis. Within a cylinder of radius $R_0$ (around the orbital axis), no shocks are formed. These results are obtained straightforwardly by projecting the wind velocity onto planes parallel to the orbital plane.

\section{Physical and numerical setup}\label{Sec: Numerical simulations}
\subsection{Numerical method and computational grid}
We have computed 3D hydrodynamical simulations using the Guacho code, which is described by \citet{Esquivel2009} and \citet{Esquivel2013}. This code solves the gasdynamic equations in a fixed Cartesian mesh, using a second-order finite volume method with HLLC fluxes \citep{Toro1994}, and a piecewise linear reconstruction of the variables at the cell interfaces with a minmod slope limiter. A radiative cooling function based on the one presented in \citet{2007MNRAS.376..861K} and \citet{Rivera2019} for lower temperatures, and the cooling function of \citet{Raga2004} for higher temperatures are included in our numerical models.

Some of our simulations have $1500\times1500\times200$ pixels
and a physical size of $0.06\times0.06\times0.008$~pc along the $(x,y,z)$-axes (see Table~\ref{table:param-second}).
The orbit lies on an $xy$-plane, and is close to one of the limits of the computational domain along the $z$-axis. Outflow conditions are applied on all of the boundaries of the computational grid. 
In order to study the full 3D morphology of the spiral shock,
we also ran a few simulations with a larger extension along the $z$-axis: with $1500\times1500\times750$ cells and a physical size of $0.06\times0.06\times0.03$ pc along $(x,y,z)$ (see Table~\ref{table:param-second}). From our simulations, we can reconstruct the full 3D flow by carrying out an appropriate reflection of the flow with respect to the orbital plane.

\subsection{The parameters of the binary system and the surrounding environment}

We compute a set of 3D gasdynamic simulations to explore
scenarios for modelling the spiral shock in the wind from a binary system considering:
\begin{itemize}
\item a single, isotropic wind from an AGB star in orbit around a wind-less   companion,
\item two winds from AGB stars that form a binary system.
\end{itemize}
In both cases, we restrict our study to the case of circular orbits, and use flow parameters appropriate for the AFGL~3068 system.

The characteristics of the binary system are taken from
the work of \citealt{Kim2012a}, who modeled AFGL~3068.
The two AGB stars have $M_1$=3.5 (for the primary component) and $M_2$=3.1 $M_{\odot}$ masses, and are in a circular
orbit. The separation between the two stars is of 166 AU
and the orbital period is $P=830$ years\footnote{These are similar values to   those reported by \citet{Mauron2006} and \citet{Kim2012a} for AFGL~3068}. All of our simulations share these parameters for the binary stellar wind source.

In all of our models the binary is initially surrounded
by a homogeneous environment with 100 cm$^{-3}$ number density and 30~K temperature. This environment is eventually flushed out of the computational grid by the wind (or winds) from the binary source. Also, in all of our simulations the stellar winds have an initial temperature of 50~K, which is fixed at the outer boundary of the wind imposition spheres (see below).


\subsection{Models with a single wind and with two winds}

In order to carry out a comparison with the analytical model of Section 2, we first compute a simulation of a single stellar wind source in a circular orbit with a windless companion (model \textbf{M0}, see Table~\ref{table:param-second}). The wind has a $\dot{M_1}= 4.0\times 10^{-5}$ $M_{\odot}$yr$^{-1}$ mass loss rate and a $v_{w,1}=12.2$ km s$^{-1}$ terminal velocity, which correspond to the powerful and fast AGB winds of \citet{Hofner2018}. These parameters are similar to those used in previous papers (see e.g. \citealt{Kim2012a,Kim2012b,Decin2015,Saladino2019a,Saladino2019b}), and represent the ``superwind'' stage that marks the
  transition from the AGB to post-AGB phase (see, e.g.,
  \citealt{Lagadec2008,Goldman2017}) rather than a more typical AGB wind. In this way, we study systems similar to those reported in the aforementioned papers. The stellar  wind is initially (and also at all times) imposed within a sphere of radius $R_w=33.0$ AU around the position of the star (see Table~\ref{table:param-second}).

We have also computed a set of models of binaries in which the two stars eject initially isotropic winds. In these models we impose the orbital parameters described in section 3.2, and the wind from the primary star in all cases has a $\dot{M_1}= 4.0\times 10^{-5}$ $M_{\odot}$yr$^{-1}$
mass loss rate and a $v_{w,1}=12.2$ km s$^{-1}$ wind velocity. We explore a set of $\dot{M_2},\,v_{w2}$ (mass loss rate, wind velocity) values for the wind from the secondary binary companion (see Table~\ref{table:param-second}).

Following the work of \citet{Canto1996} we define the wind
ram-pressure ratio, $\eta=\dot{M_2}v_{w2}/\dot{M_1}v_{w1}$\footnote{our $\eta$ parameter is the inverse of the $\beta$ parameter of \citet{Canto1996}.}; and the velocity ratio $\alpha={v_{w2}}/{v_{w1}}$ of the two winds.

Table~\ref{table:param-second} gives:
\begin{itemize}
  \item the mass loss rate and the velocity
    for the wind from the companion star,
  \item the $\eta$ and $\alpha$ factors defined above,
  \item the radius $R_w$ of the spheres (around the instantaneous positions of the stars) in which the winds are imposed,
  \item the distance $R_s$ of the two-wind interaction stagnation point measured from the position of the primary star (calculated from the condition of on-axis ram-pressure balance between two winds, see \citealt{Canto1996}),
  \item the physical size and the resolution of the computational domain.
\end{itemize}
We can see that in all cases the stagnation region lies clearly outside the spheres in which the winds are imposed.
It is important to note that despite the fact that some of our models (e.g. models \textbf{M1a}, \textbf{M1c} and \textbf{M3}) have secondary wind velocities which are low even for an AGB star, the $\dot{M_2}v_{w2}$ term are not necessarily so low. \citet{Kim2012b} reported that the rates of mass loss from AGB stars are in between $10^{-7}-10^{-4}$ $M_{\odot}$ yr$^{-1}$, on the other hand, \citet{Ramstedt2014} estimated the mass-loss rate from Mira A a few times $10^{-7}$ $M_{\odot}$ yr$^{-1}$. They expected a low wind velocity of approximately 5 km s$^{-1}$. However our main intention on the selection of the parameter for the companion star was just as a parameter study rather than based in observations.

From Table~\ref{table:param-second} we see that the ram-pressure (or momentum flux) of the wind from the primary star dominates over the wind from the secondary (i.e., $\eta<1$, see above) for all models except model \textbf{M4}, which has $\eta=1$. Also, the velocity of the primary wind is greater than the velocity of the wind from the secondary star (i.e., $\alpha<1$), except for models \textbf{M1b}, \textbf{M2}, and \textbf{M4}, which have $\alpha=1,\, 2.05,\, 1$ respectively.

\begin{table*}
\begin{center}
\caption{Parameters for the companion star in all models.}
\label{table:param-second}
\begin{threeparttable}
\begin{tabular}{|l|l|l|l|l|l|l|l|}
\hline
Model & \textbf{M0} & \textbf{M1a} & \textbf{M1b} & \textbf{M1c} & \textbf{M2} & \textbf{M3} & \textbf{M4} \\
$\dot{M_2}$ ($M_{\odot}$yr$^{-1}$) & $\ldots$ &  2.71$\times 10^{-5}$ & 4.0$\times 10^{-6}$ & 4.0$\times 10^{-5}$ & 9.76$\times 10^{-6}$ & 2.71$\times 10^{-5}$ & 4.0$\times 10^{-5}$\\
$v_{w2}$ km s$^{-1}$ & $\ldots$ &  1.8 &  12.2 &  1.22 & 25 & 2.7 & 12.2 \\
$\eta$ & $\ldots$ & 0.1 & 0.1 & 0.1 & 0.5 & 0.15 & 1.0 \\
$\alpha$ & $\ldots$ &  0.15 & 1.0 & 0.1 & 2.05 & 0.22 & 1.0 \\
$R_w$ (AU) & 33.0 & 33.0 & 33.0 & 33.0 & 33.0 & 33.0 & 33.0 \\
$R_s$ (AU) & $\ldots$ & 126.12 & 126.12 & 126.12 & 97.24 & 119.65 & 83 \\
pixels\tnote{a} & pix1 & pix1 & pix2 & pix1 & pix2 & pix2 & pix2 \\
size\tnote{b} (pc) & sz1 & sz1 & sz2 & sz1 & sz2 & sz2 & sz2 \\
\emph{eddies} & no & yes & no & yes & no & yes & no\\
\hline
\end{tabular}
\begin{tablenotes}
\item[a] Refers to the number of points by side of the simulation: pix1= $1500\times 1500\times 750$; pix2=$1500\times 1500\times 200$.
\item[b] Refers to the physical size of the computational domain:
  sz1=$0.06\times 0.06\times 0.03$~pc; sz2=$0.06\times 0.06\times 0.008$~pc.
\end{tablenotes}
\end{threeparttable}
\end{center}
\end{table*}

\section{Results}\label{Sec: Results}

\subsection{Model with a single wind}

\begin{figure}
\centering
\includegraphics[width=\columnwidth]{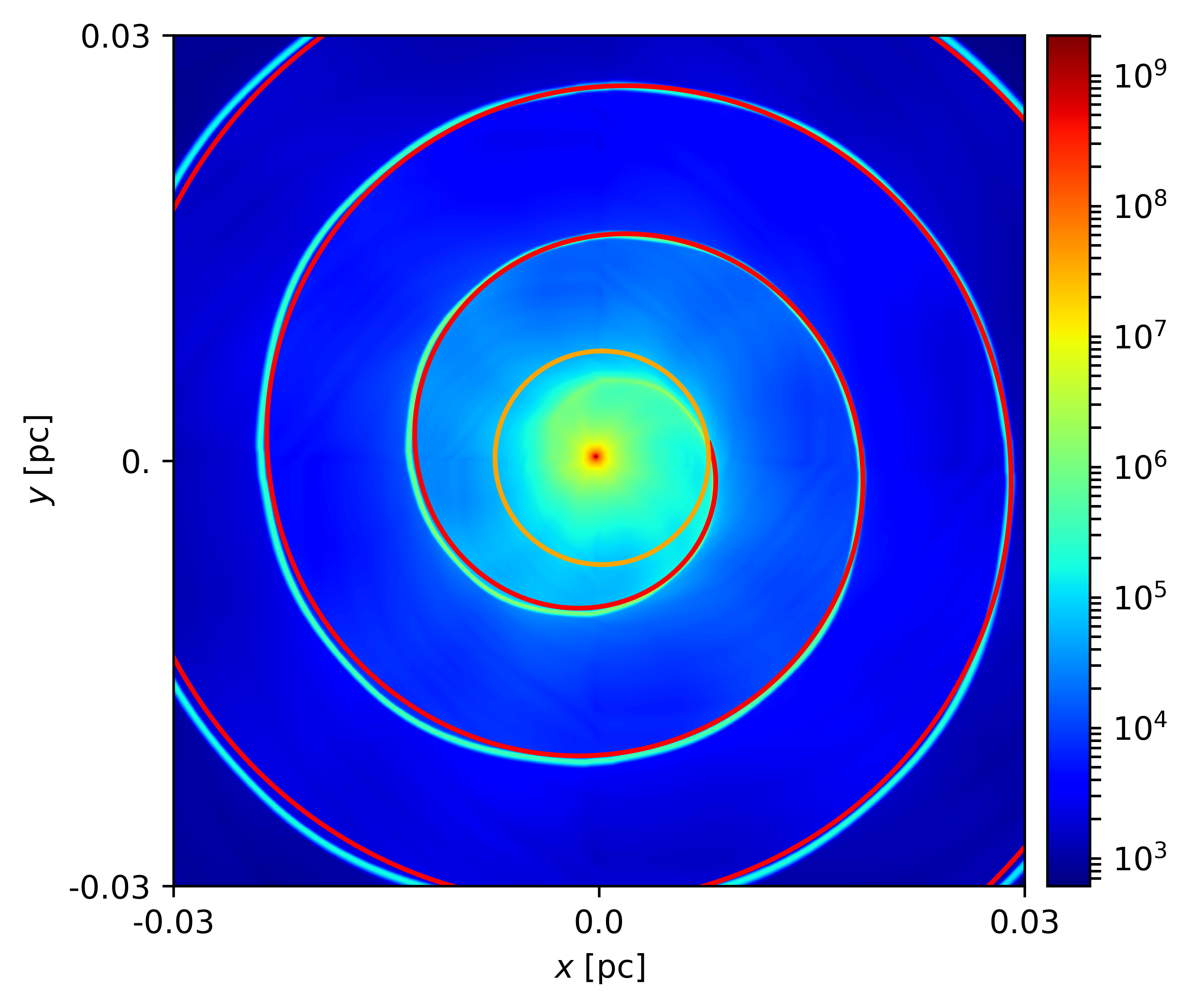}
\caption{The number density stratification on the orbital plane of the single wind model (model \textbf{M0} of Table~\ref{table:param-second}) after a 4000~yr time-integration is shown with the logarithmic colour scale given (in cm$^{-3}$) by the bar on the right. The red curve shows the spiral structure predicted by the analytic model, and the orange circle shows the starting radius at which this shock structure begins. The axes are labeled in pc.}
\label{fig:Fig1}
\end{figure}

The model with a single stellar wind (model \textbf{M0} of Table~\ref{table:param-second}) shows the typical structure of a spiral shock. Figure~\ref{fig:Fig1} shows the density structure on the orbital plane after a time-integration of 4000~yr together with the spiral shock structure predicted from the analytical model (see section 2) for the same evolutionary time (shown with the solid, red curve). We find a good agreement between the spiral structures of the analytic model and the numerical simulations. We have also drawn an orange circle with the radius $R_0$ (given by
equation~\ref{eq:startrad}) at which the analytic model predicts the beginning of the shock structure.

\begin{figure}
\begin{center}
\includegraphics[width=\columnwidth]{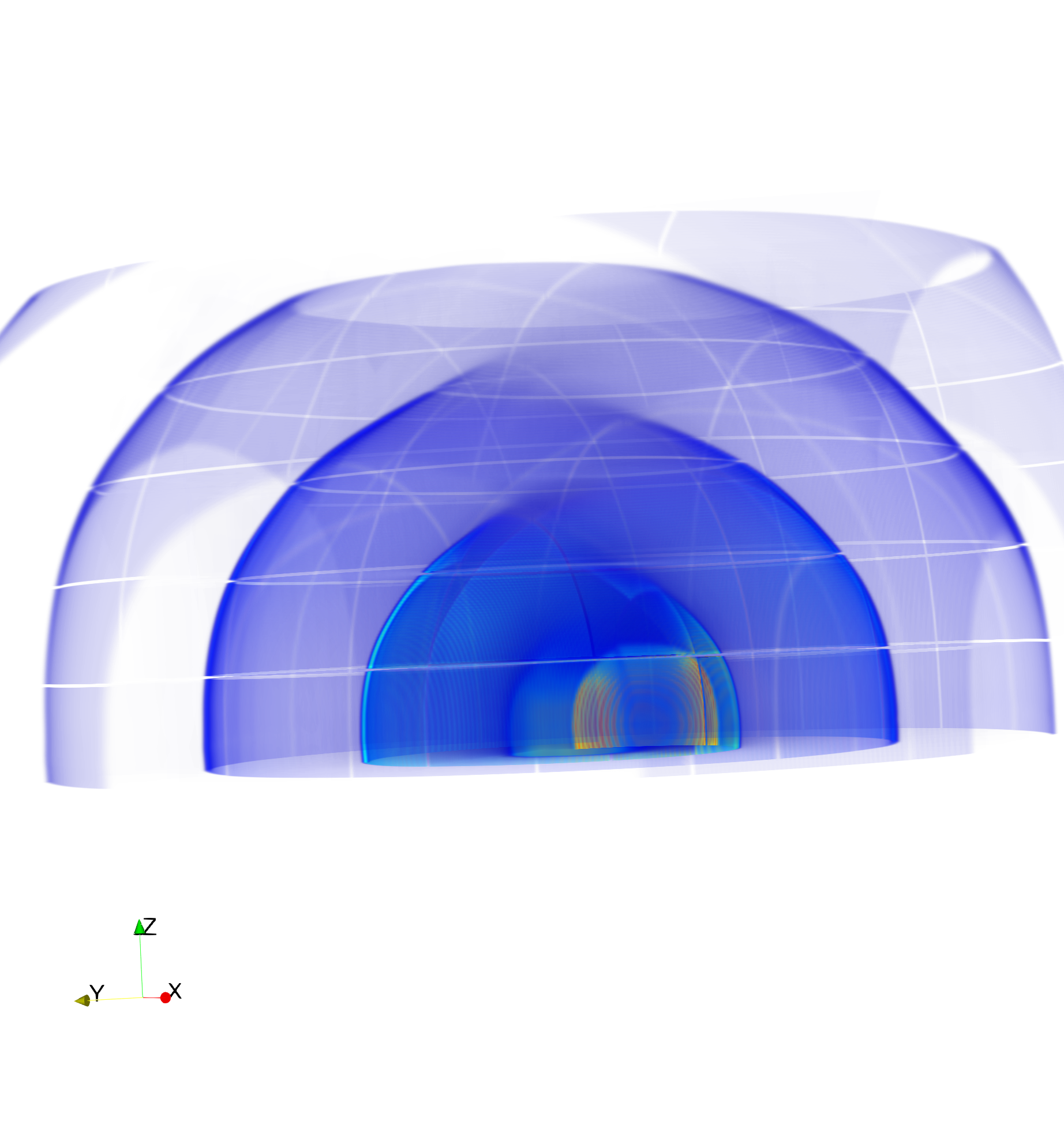}
\caption{3D density stratification of model \textbf{M0} at a
  4000 yr time. We observe a concentric spherical shell morphology as a result of the stellar wind self-interaction due to the orbital velocity. We note that, as predicted by the analytical model, there is a free-wind region around the orbital axis.}
\label{fig:Fig2}
\end{center} 
\end{figure}

The 3D morphology of the shocks can be seen in the volume rendition of the 3D density structure of model \textbf{M0}, shown in Figure~\ref{fig:Fig2}. We see that the density structure
has the arc-like morphologies and the shock-free cylindrical
region (around the orbital axis) predicted by the analytic
model described in section 2. The flattened part in the outermost region on the top of Figure 2 corresponds to the top, outflow boundary of the computational domain.

\subsection{Models with two winds}

\begin{figure*}
\begin{center}
\includegraphics{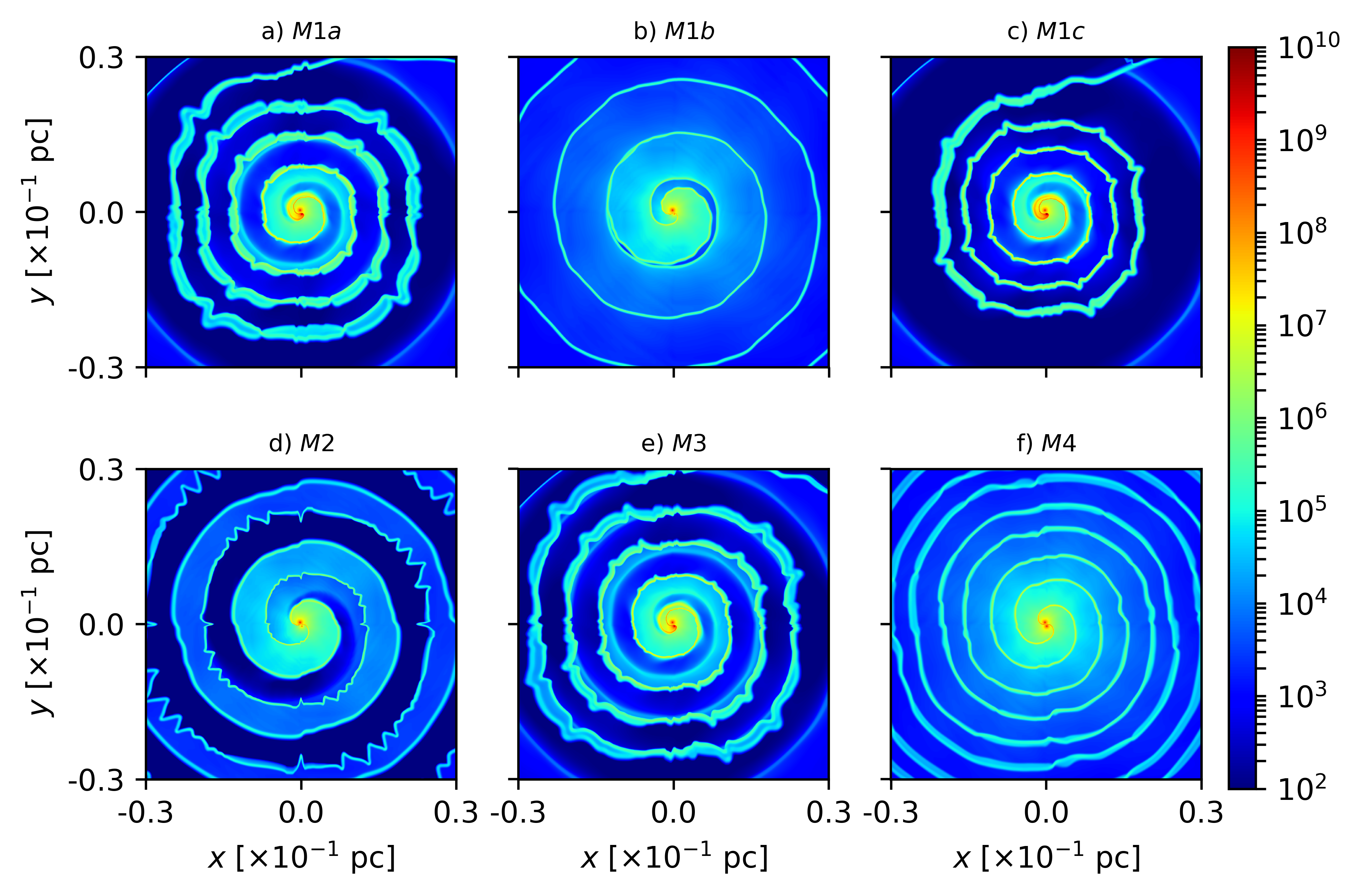}
\caption{Density stratification in the orbital plane
  for models \textbf{M1a}-\textbf{M4} at a time of 4000 yr.
  The top row shows models \textbf{M1a}, \textbf{M1b} and \textbf{M1c} (panels a-c), and the bottom row shows models \textbf{M2}, \textbf{M3} and \textbf{M4} (panels d-f), which have the parameters given in Table~1. The number densities are given (in cm$^{-3}$) by the bar on the right, and the axes are labeled in units of 0.1~pc. Model \textbf{M4} (with two identical winds, see Table~\ref{table:param-second}) shows a remarkable twin spiral structure. Models \textbf{M1a}, \textbf{M1c}
  and \textbf{M3}, with wind velocities differing by factors of $\sim 5\to 10$ (see Table~1) develop complex eddies (see the text).}
\label{fig:Fig3}
\end{center} 
\end{figure*}

In Figure~\ref{fig:Fig3}, we show the density stratification for models \textbf{M1a}-\textbf{M4} at a 4000 yr time. We note that in all cases a spiral shock structure has developed.
This spiral structure can be fit by a single or a double Archimedean spiral with arms moving outwards at a velocity intermediate between the velocities of the two stellar winds.

Models \textbf{M1a}, \textbf{M1b}, and \textbf{M1c} (panels a, b and c in  Figure~\ref{fig:Fig3}), which share an $\eta=0.1$ secondary to primary ram-pressure (or momentum flux) ratio (see Table~\ref{table:param-second}) show qualitatively different structures: models \textbf{M1a} and \textbf{M1c} develop a strongly perturbed structure with a number of complex eddies, while model \textbf{M1b} shows relatively low amplitude ``thin shell instabilities'' along the spiral structure.

Such low amplitude thin shell instabilities in the spiral structure are obtained when the two winds have identical velocities (models \textbf{M1b} and \textbf{M4}, which have $\alpha=1$, see Table~\ref{table:param-second}).
Larger amplitude perturbations are seen in the \textbf{M2} model, (which has a wind velocity ratio $\alpha=2.05$,
see Table~\ref{table:param-second} and Figure~\ref{fig:Fig3}). The models with larger velocity differences between the primary and secondary winds (models \textbf{M1a}, \textbf{M1c} and \textbf{M3}, with $\alpha=0.15$, 0.1 and 0.22, respectively) show the development of large amplitude perturbations in the resulting
spiral structures (this is indicated in the last line
of Table~\ref{table:param-second}).

The thin shell instability in two-wind interactions was first   studied by \citet{Stevens1992}. This paper considers
  the interaction of two massive stellar winds, but not considering the orbital motion of the stars. The effect of the orbital motion has been studied in other papers, e.g. in \citet{vanMarle2011} and \citet{Lamberts2012}. According to \citet{Stevens1992} the thin shell instabilities develop in cases where cooling is important (i.e., the higly radiative case), while if the cooling is negligible (the quasi-adiabatic case), the thin shell instability is not active.

\begin{figure}
\begin{center}
\includegraphics[width=\columnwidth]{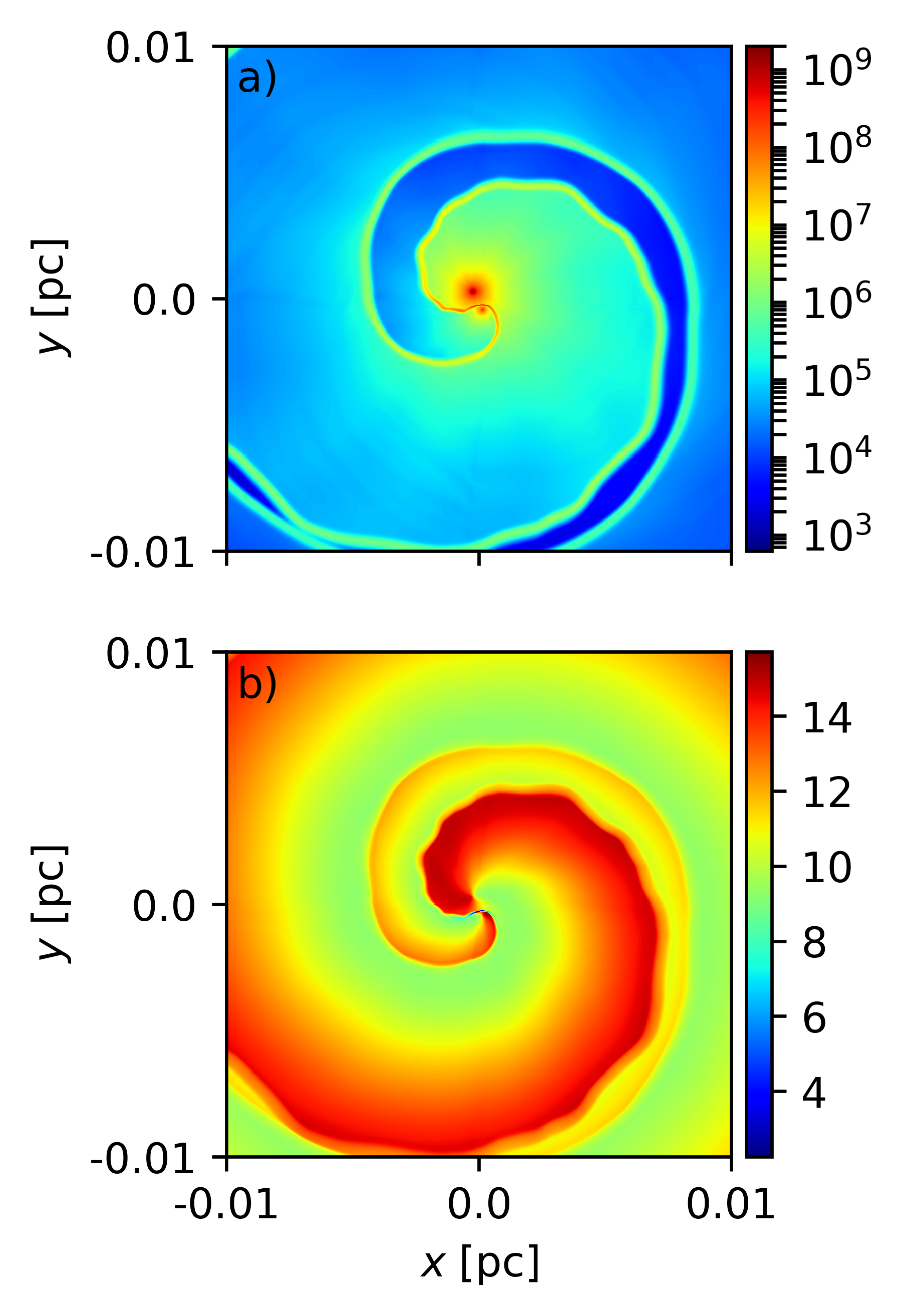}
\caption{Zoom into the region close to the binary wind source of model \textbf{M1b}. The density (panel a, with values in cm$^{-3}$) and velocity modulus (panel b, with bar values in kms$^{-1}$) orbital plane stratifications are shown. The perturbations in the spirals due to the thin shell instabilities are clearly seen.}
\label{fig:Fig4}
\end{center}
\end{figure}

The thin shell instability is active in our models, since in all cases we are in the highly radiative regime (we have post-shock cooling/dynamical time ratios $<10^{-5}$). In our models with equal velocity winds, we see relatively small amplitude inhomogeneities which probably arise from the thin shell instability (see the models \textbf{M1b} and \textbf{M4} in Figure 3). This is clearly seen in Figure~\ref{fig:Fig4}, which shows a zoom into the central region of model \textbf{M1b} (the top panel showing the density, and the
    bottom panel the velocity modulus stratification on the orbital plane).

In the remaining models (\textbf{M1a}, \textbf{M1c}, and  \textbf{M3} models), the thin shell instability is probably also present (because the models are in the highly radiative regime). However, the fact that the perturbations grow to considerably larger amplitudes is likely to be due to a combination of the thin shell and the Kelvin-Helmholtz (K-H) instability resulting from the velocity difference between the two winds. The large amplitudes in our larger wind velocity ratio models (models \textbf{M1a} and \textbf{M1c}) indicates a dominance of the K-H instability.

In all of the two-wind models, close to the binary we have a double spiral structure, corresponding to the intersection with the orbital plane of the wings of the two-wind interaction bow shock. These two spirals merge into a single spiral structure at larger distances from the source in all models except \textbf{M4} (which has two identical winds, see Table~\ref{table:param-second}).

\begin{figure}
\begin{center}
\includegraphics[width=\columnwidth]{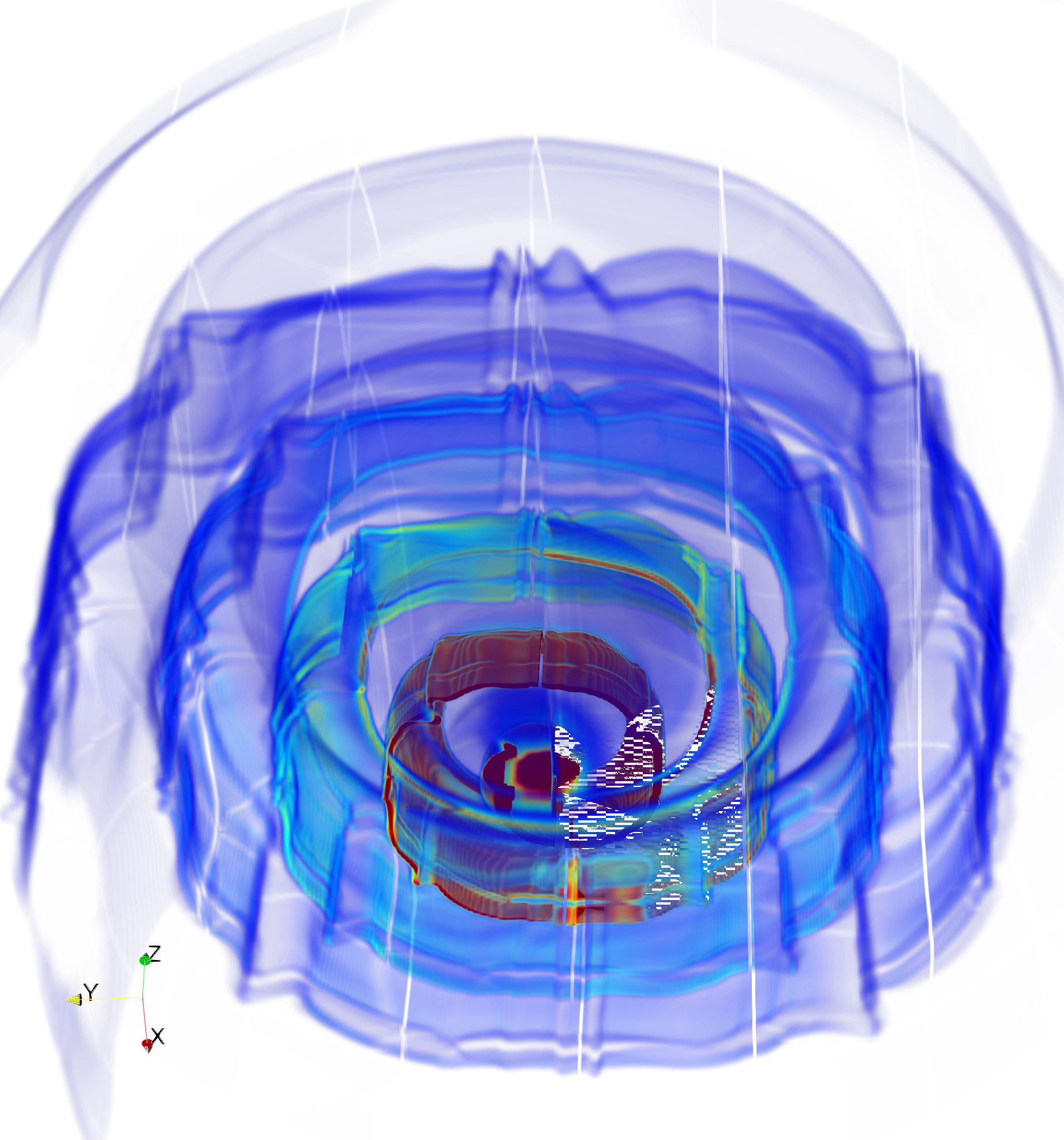}
\caption{3D density stratification for model \textbf{M1a} at a time of 4000~yr. We see that the perturbed \emph{eddies} are confined to a region close to the orbital plane.}
\label{fig:Fig5}
\end{center}
\end{figure}

In Figure~\ref{fig:Fig5} we show a 3D rendition of the density stratification of model \textbf{M1a}. We see that the flow with perturbed eddies is confined to a region close to the orbital plane, and that an unperturbed flow with orderly shocks is found closer to the orbital axis.

\begin{figure}
\begin{center}
\includegraphics[width=\columnwidth]{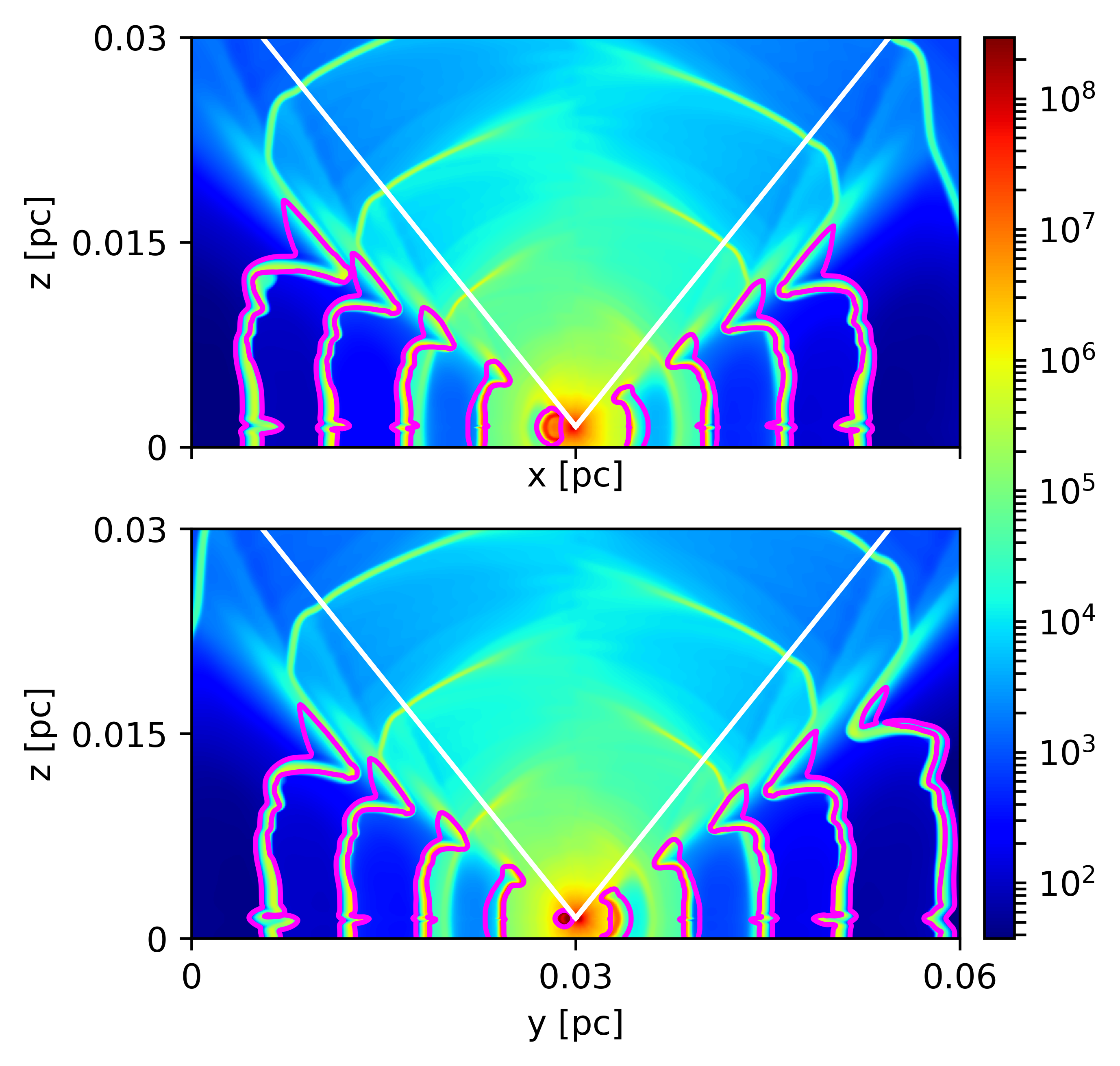}
\caption{Density stratification of model \textbf{M1a} for two slices including the orbital axis. Top: $xz$ slice, and bottom: $yz$ slice. The magenta contour encloses the regions occupied by the weaker wind from the secondary star (traced with a passive scalar) and the straight, white line shows the asymptotic opening angle of the two-wind interaction bow shock (see the text).}
\label{fig:Fig6}
\end{center}
\end{figure}

\begin{figure}
\begin{center}
\includegraphics[width=\columnwidth]{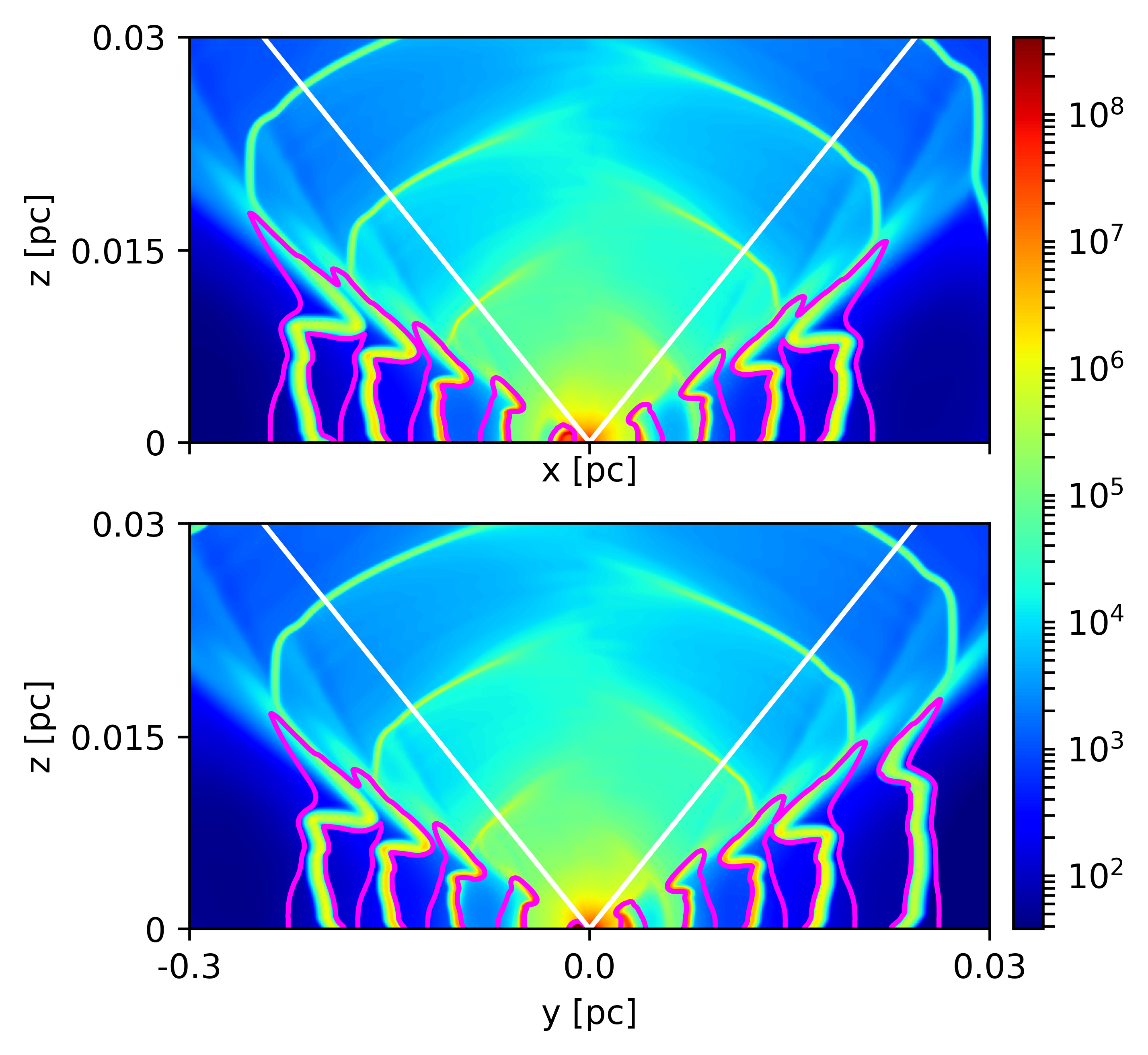}
\caption{Same as Figure~\ref{fig:Fig5} but for model \textbf{M1c}.}
\label{fig:Fig7}
\end{center} 
\end{figure}

This behaviour is also clearly seen in Figures~\ref{fig:Fig6} and \ref{fig:Fig7}, which show the density stratifications of
models \textbf{M1a} and \textbf{M1c} on the $xz$- and $yz$-planes (which include the orbital axis). In these
Figures we also show a contour (in magenta) that encloses the regions occupied by the wind from the secondary star, which we have traced with a passive scalar. It is clear that (as expected), the lower momentum rate, secondary wind is confined to a limited region centered on the orbital plane, and that the stronger wind from the primary source solely occupies an approximately conical region around the orbital axis.

In Figures~\ref{fig:Fig6} and \ref{fig:Fig7} we also show
(with a white, straight line) the asymptotic opening angle of the two-wind interaction bow shock wings predicted from the model of \cite{Canto1996} (see their equation 28). It is
clear that the region between this straight line and the polar axis is free from the perturbations resulting from the two-wind interaction, having a shock structure similar to the one of the single wind model \textbf{M0} (see Table~\ref{table:param-second} and Figure \ref{fig:Fig2}).

\subsection{Predicted dust emission}

In order to have predictions that can be directly compared with AFGL 3068, in which the spiral structure is seen in the
dust-scattered stellar continuum (see \citealt{Mauron2006}), we have calculated predicted scattered light maps from some of our models. These maps are calculated in the most simple possible way, assuming a single, isotropic scattering, no dust extinction, and perfect dust/gas dynamical coupling with a constant dust-to-gas ratio throughout all the simulation box. The intensity maps that we compute do not include the thermal dust emission, which is important in the IR, but not at optical wavelengths.

\begin{figure}
\begin{center}
\includegraphics[width=\columnwidth]{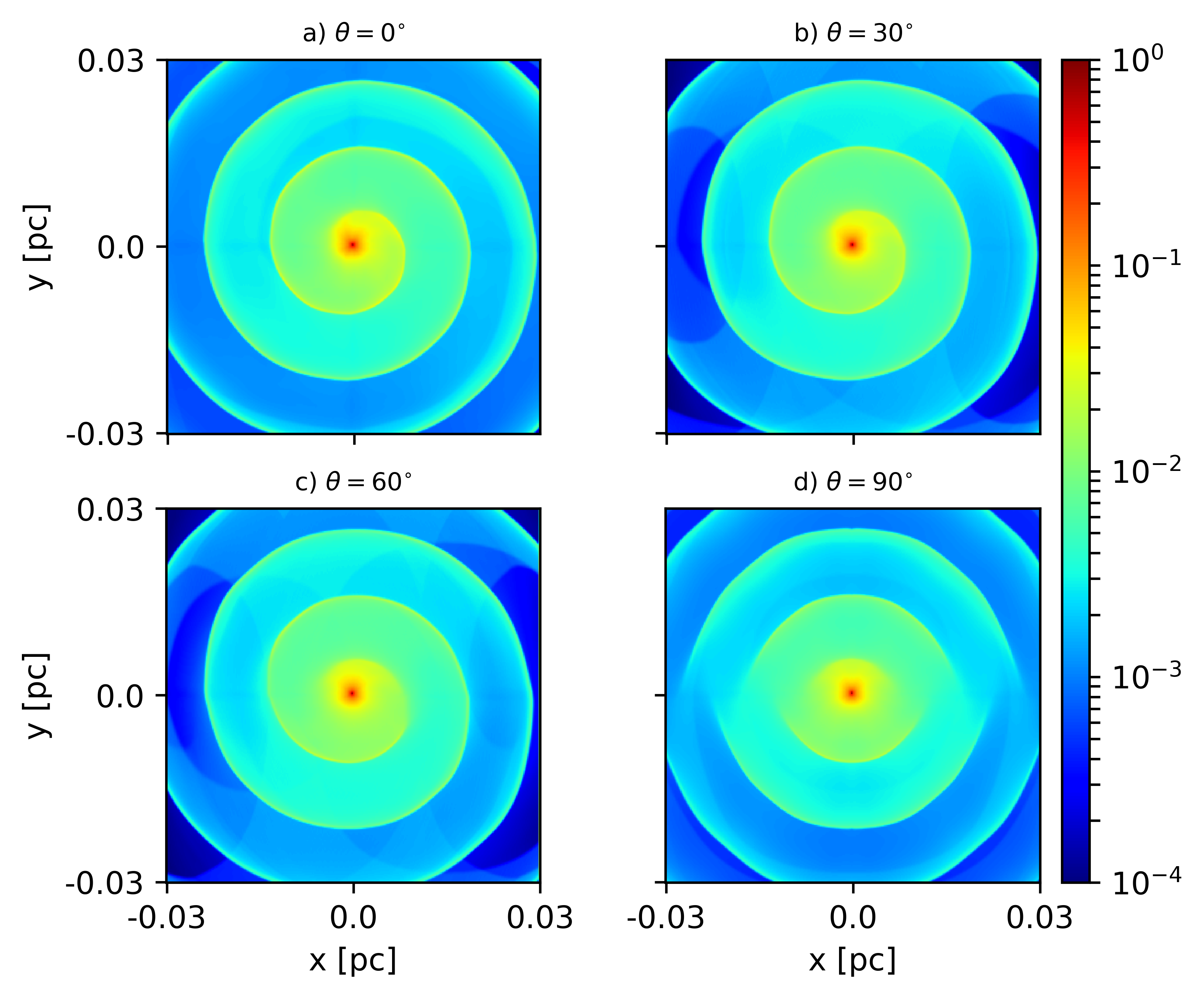}
\caption{Dust scattered stellar continuum map for model \textbf{M0}, calculated for $\theta=0$, 30, 60 and $90^\circ$ orientations between the orbital axis and the line of sight. In the $\theta\neq 0$ maps, the projection of the orbital axis on the plane of the sky is parallel to the ordinate.}
\label{fig:Fig8} 
\end{center} 
\end{figure}

\begin{figure}
\begin{center}
\includegraphics[width=\columnwidth]{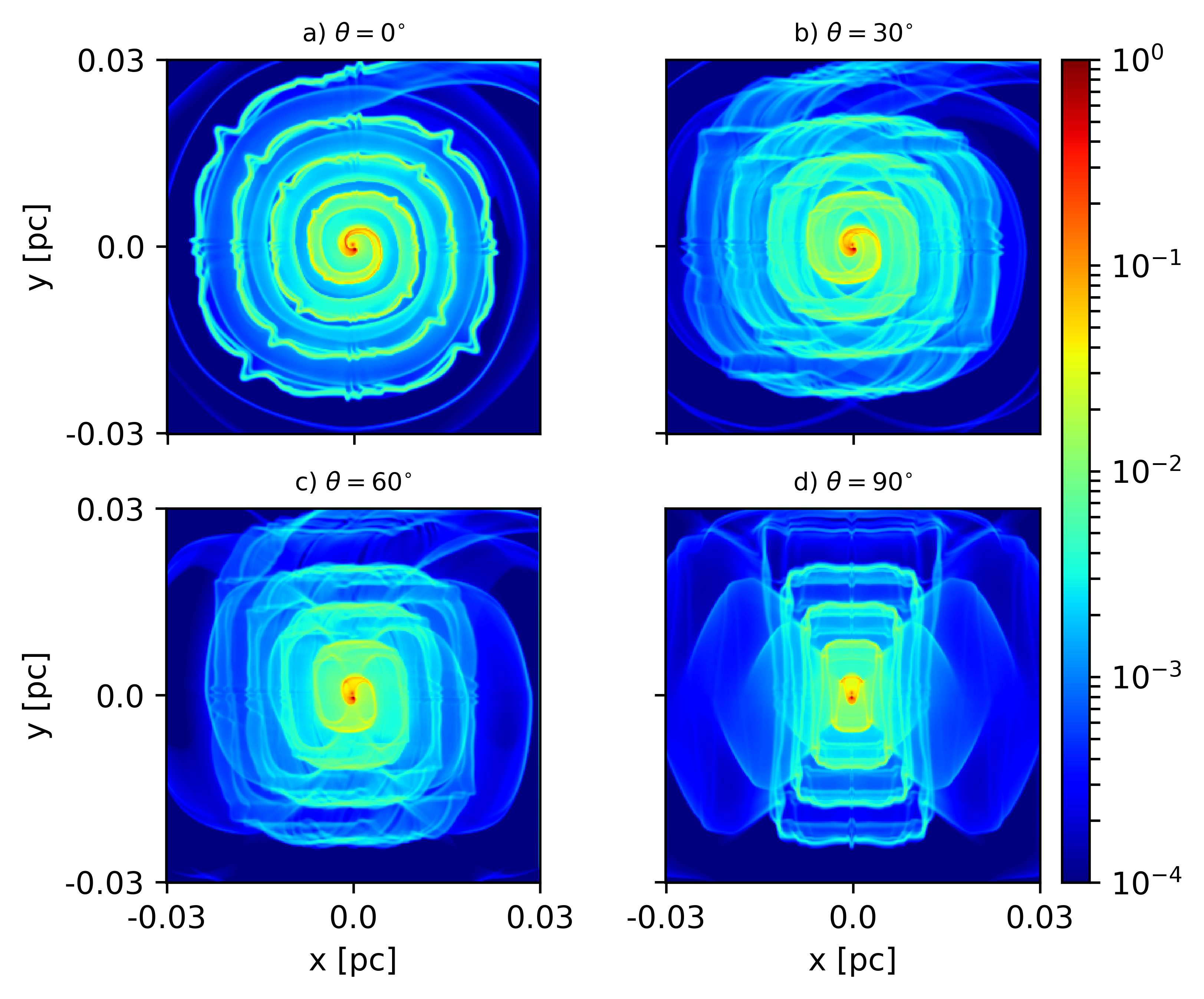}
\caption{Same as Figure~\ref{fig:Fig7} but for model \textbf{M1a}.}
\label{fig:Fig9}
\end{center} 
\end{figure}

\begin{figure}
\begin{center}
\includegraphics[width=\columnwidth]{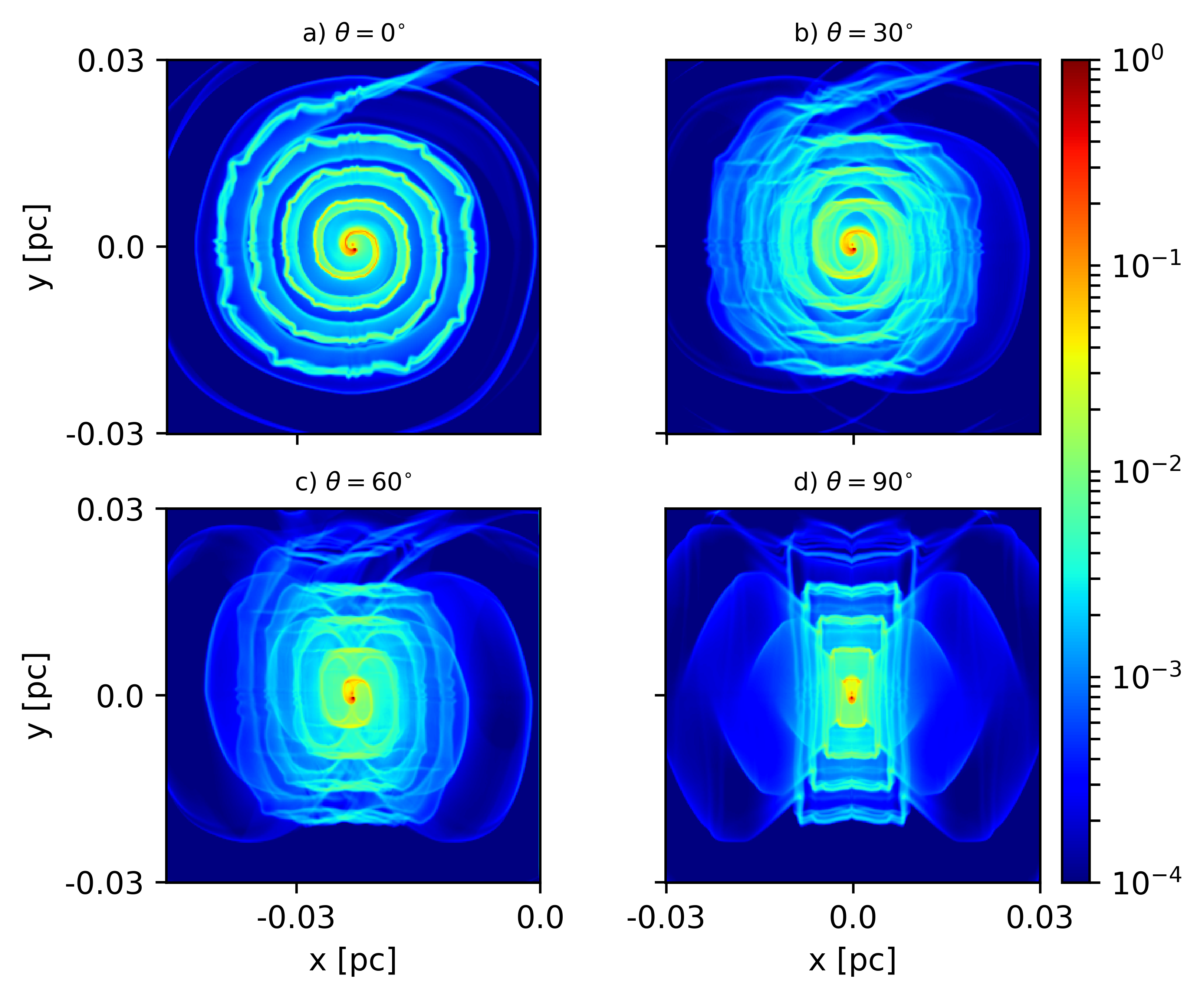}
\caption{Same as Figure~\ref{fig:Fig7} but for model \textbf{M1c}.}
\label{fig:Fig10}
\end{center} 
\end{figure}

In Figures \ref{fig:Fig8}-\ref{fig:Fig10}
we show the scattered continuum maps predicted
from models \textbf{M0}, \textbf{M1a} and \textbf{M1c}, respectively. Each figure shows the intensity maps predicted for four different values of the angle $\theta$ between the orbital axis and the line of sight ($\theta=0$, 30, 60 and $90^\circ$).

For the single wind \textbf{M0} model (see Figure~\ref{fig:Fig8}), an orderly spiral structure is seen for all of the chosen orientations except for $\theta=90^\circ$. In this case with the orbital axis on the plane of the sky, a morphology of successive, separate arcs is obtained, with clear breaks in the region close to the orbital axis.

For models \textbf{M1a} and \textbf{M1c}, a clumpy spiral is observed for the $\theta=0$ (observer along the orbital axis) and $\theta=30^\circ$ orientations, see Figures \ref{fig:Fig9} and \ref{fig:Fig10}. For both models, the $\theta=60^\circ$ maps develop morphologies of a spiral with a squarish shape, and the $\theta=90^\circ$ maps show a stepped, biconical structure aligned with the orbital axis.

\section{Summary}

We computed 3D gasdynamical simulations of the by now ``classical problem'' of the wind from a stellar source (assumed to be a red giant) in a binary orbit, and also of a binary with two giants with powerful winds. For the case of a single wind, we confirm that both the numerical and analytical model resemble the observations of AFGL~3068 of \citet{Mauron2006} (see also \citealt{Raga2011}). The predictions of the dust scattered stellar continuum (see Figure~\ref{fig:Fig8}) show that a spiral similar to the observed one is obtained for most orientation angles,
except for cases in which the orbital axis lies less than $\sim 20^\circ$ from the plane of the sky, in which the shock-free zone close to the orbital axis shows up as breaks in the observed spiral.

For all of the two-wind models, we find a double spiral morphology on the orbital plane (see Figure 3), corresponding to the two-wind interaction bow shock wings. These two spirals merge into a single spiral structure at larger distances from the source, except for model \textbf{M4} (with two identical winds, see Table~\ref{table:param-second}) in which the double spiral structure is preserved (see Figure \ref{fig:Fig3}).

The presence of a double spiral structure (at least, close to
the binary wind source) therefore appears to be the signature
of the existence of a dynamically important wind from
the secondary star in an observed system. AFGL~3068
(\citealt{Mauron2006}) does not appear to show such a double
spiral structure.

An interesting feature of our simulations is that while models with two winds of similar velocities (i.e., models \textbf{M1b}, and \textbf{M4}, with wind velocity ratios
$\alpha=1.0$ wind velocity ratios, see Table~\ref{table:param-second}) show spirals with only low amplitude perturbations (due to the thin shell instabilty),
the models with larger wind velocity contrasts (i.e., models \textbf{M1a}, \textbf{M1c} and \textbf{M3}, see Table~\ref{table:param-second}) show spirals that break into a number of small-scale ``eddies'' associated with Kelvin-Helmholtz instabilities. The signature of the presence
of a large velocity contrast (in our models, a contrast
of $\sim 2$ or greater) between the winds from the two
stars in the binaries therefore is the formation of clump-like features along the spiral structure.

Clump-like structures have been observed in some spiral patterns such as in the CO molecular line emission of R~Sculptoris presented by \citet{Maercker2012}, the HC 3 N molecular line emission in CIT~6 by \citet{Kim2013}, 
the CO molecular line emission by \citet{Guelin2018} of
IRC~+10~216; and the CO, molecular line emission detected by
\citet{Homan2018} around the AGB star EP~Aquarii. According to our models, these structures could be tentatively interpreted as evidence for the presence of two winds with
a large wind velocity contrast. 

The density structures perpendicular to the orbital plane of
the two-wind models differ strongly from the single wind
model. The two-wind models have a region close to the orbital
plane which is occupied by both stellar winds (see Figures \ref{fig:Fig6} and \ref{fig:Fig7}), and an approximately
conical region around the orbital axis which is solely occupied by the stronger wind (this result does not hold for model \textbf{M4}, with two identical stellar winds).

This division into two separate flow regions (an equatorial region with mixed winds, and a polar region with only the stronger wind) leads to complex morphologies in the predicted dust scattered continuum maps. The maps predicted from models \textbf{M1a} and \textbf{M1c} (Figures \ref{fig:Fig9} and \ref{fig:Fig10}, respectively) show strong deviations from an Archimedean spiral for orientation angles $\theta=60$ and $90^\circ$.

Quite strikingly, the maps predicted from these models
(\textbf{M1a} and \textbf{M1c}, see Figures \ref{fig:Fig9} and \ref{fig:Fig10}) for $\theta=90^\circ$ show a stepped, bipolar conical structure (centered on the orbital axis). This kind of structure might be appropriate for modelling Planetary Nebulae such as HD~44179 (the ``red rectangle'', see \citealt{Cohen2004})
or Hubble 12 \citep{Vaytet2009}.

It is important to note that when PNe are formed the fast wind of the post-AGB star starts sweeping up the AGB superwind. Thus, the morphology of PNe will be determined by the structure of the post-AGB wind (such as the ones obtained in our simulations) only in the protoplanetary nebula (PPN) regime or in the early evolution of the PN. In this early evolution, the high velocity wind from the post-AGB central source will be confined to a limited circumstellar region, and the outer part of the nebula will be filled in with a partially photoionized remnant of the AGB wind. In order to try to model the initial stages of PNe, it will be interesting to carry out simulations of   single or two-wind interaction structures from a binary in which one of the stars begins to have a fast post-AGB wind and starts to emit photoionizing radiation.

The models that we have computed represent a small sampling
of the parameter space of the problem of a binary with two
stellar wind sources, restricted to the case of two giant
stars in circular orbits. Clearly, a broader range of
situations remain to be explored, including the case of elliptical orbits (which has been studied for the case of a single wind from an orbiting source) and other combinations of winds (e.g., a fast wind from a compact, hot star and a giant star wind).

We conclude by noting that these binary wind interactions are not only important for modelling objects in which we see spiral structures (see the discussion of section 1), but also for modelling planetary nebulae and supernova remnants which
expand into circumstellar environments shaped by a previous
binary wind interaction.

\section*{Acknowledgements}
This work was supported by DGAPA (UNAM) grant IG100218. A.C.-R. acknowledges support from a DGAPA-UNAM postdoctoral fellowship 
and the support provided by PAPIIT: IA103121. A.C.-R and A.R.-G. acknowledge the resources provided by the Miztli supercomputer through the project LANCAD-UNAM-DGTIC-408. P.R.R.-O. acknowledges funding from the European Research
Council (ERC) under the European Union's Horizon 2020
research and innovation program, for the Project ``The
Dawn of Organic Chemistry'' (DOC), grant agreement 
No. 741002. We thank an anonymous referee for clear comments
that lead to a substantial rewriting of the paper.

\section*{Data Availability}
The data underlying this article will be shared on reasonable request to the corresponding author.
 



\bibliographystyle{mnras}
\bibliography{references} 








\bsp	
\label{lastpage}
\end{document}